\documentclass[sigconf]{acmart}

\usepackage{amsmath}
\usepackage{graphicx}
\usepackage{booktabs}
\usepackage{multirow}
\usepackage{hyperref}

\setcopyright{acmlicensed}
\copyrightyear{2026}
\acmYear{2026}
\acmConference[Conference '26]{Proceedings of Conference '26}{June 2026}{Location}
\acmBooktitle{Proceedings of Conference '26, June 2026, Location}
\acmDOI{10.1145/XXXXXX.XXXXXX}
\acmISBN{978-1-4503-XXXX-X/26/06}
\begin{document}

\title{Decoupled Residual Quantization for Robust Semantic IDs in Recommendation}

\author{\texorpdfstring{Xuesi Wang\textsuperscript{$\dagger$}, Junjie Wang, Ziliang Wang, Weijie Bian\textsuperscript{*}, Guanxing Zhang}{Xuesi Wang, Junjie Wang, Ziliang Wang, Weijie Bian, Guanxing Zhang}}
\renewcommand{\shortauthors}{Xuesi Wang et al.}
\affiliation{
  \institution{Shopee}
  \country{}
}

\begin{abstract}
Semantic IDs represent items as shared discrete token sequences and have become a practical tool for recommendation and retrieval. Yet it remains difficult to tell why a tokenizer fails: poor quality may come from codebook underutilization, unstable decision boundaries, or geometric distortion of the embedding space. This paper develops a quantitative framework for diagnosing these failures through expected codeword overlap and effective codebook capacity. The former measures expected codeword confusion under retrieval-time perturbation, while the latter converts that confusion into an effective number of usable, well-separated codes. The framework links semantic boundary confusion to both code usage imbalance and Euclidean geometric constraints. As a proof of concept, we present Decoupled Residual Quantization (DRQ), which separates continuous geometry reconstruction from discrete distribution matching. Experiments on a large-scale industrial dataset show that Semantic ID quality is multi-objective: symbolic robustness, reconstruction fidelity, and behavior-aware soft matching each stress different aspects of a tokenizer. These downstream observations are based on one proprietary industrial dataset, so they should be read as a case study rather than a universal benchmark claim.
\end{abstract}

\ccsdesc[500]{Information systems~Recommender systems}

\keywords{Recommendation, Semantic ID, Vector Quantization, Item-to-Item Retrieval}

\maketitle
\begingroup
\renewcommand{\thefootnote}{} 
\footnotetext{%
  $\dagger$ Email: \href{mailto:wxsthu@icloud.com}{wxsthu@icloud.com}. \quad \quad 
  $^*$ Corresponding author : \href{mailto:jason.bian@shopee.com}{jason.bian@shopee.com}.%
}
\endgroup

\section{Introduction}
Modern recommendation systems increasingly rely on shared discrete item representations to support retrieval, indexing, compression, and parameter sharing across items. A common approach is to quantize high-dimensional item embeddings into discrete token sequences, widely known as Semantic IDs \cite{rajput2023recommender, onerec2024}. By replacing isolated hash IDs with structured discrete codes, Semantic IDs make it easier to transfer statistical strength across items and to build unified representations over multimodal and collaborative signals.

While the value of Semantic IDs is well established, their quality depends on how continuous item embeddings are discretized. Current tokenizers often rely on established Vector Quantization (VQ) methods such as Product Quantization (PQ) \cite{jegou2010product} and Residual Quantization (RQ), with Residual-Quantized Variational Autoencoder (RQ-VAE) \cite{zeghidour2021soundstream} serving as a representative neural implementation. When these VQ methods are applied directly to recommendation embeddings, two limitations become especially important:

First, \textbf{Index Collapse and Distribution Mismatch}. Traditional VQ methods implicitly or explicitly favor balanced code usage. Real-world item interactions, however, are highly long-tailed. When a long-tailed item distribution is forced into a uniformly allocated codebook, rare items can be absorbed by a small number of popular codewords, while many other codewords receive too few assignments or gradient updates to remain useful.

Second, \textbf{Dimensional Collapse and Geometric Mismatch}. Many quantizers organize a continuous embedding space by laying down Euclidean cells and assigning each item to the nearest cell. In PQ, the vector is split into fixed coordinate blocks and each block is quantized independently. The Cartesian product of these sub-codebooks is therefore equivalent to an axis-aligned grid in the original space \cite{jegou2010product}. In neural RQ models such as RQ-VAE, the representation is built in stages: each level quantizes the residual error left by previous levels, and the final reconstruction is the sum of codewords chosen from multiple codebooks. Geometrically, this creates a Minkowski-sum codebook, the set of all possible sums of one codeword from each residual level \cite{zeghidour2021soundstream}. These designs are efficient, but they still describe the latent space with flat Euclidean building blocks. Recommendation embeddings can be highly anisotropic or concentrated on curved manifolds. When a flat grid is imposed on such data, many possible code combinations may fall in regions where no real items live, while useful item neighborhoods are compressed into a smaller effective space.

To diagnose these problems without tying the analysis to one tokenizer implementation, we study Semantic IDs through the geometry and distribution of their code assignments. The main contributions of this paper are:
\begin{enumerate}
    \item \textbf{Theoretical Framework (Core Contribution)}: We define the Expected Overlap Rate ($O_\pi$), which measures expected codeword confusion under perturbation, and the Effective Codebook Size ($K_{\text{eff}}$), which converts this confusion into an equivalent number of robust codewords. The analysis decomposes Semantic ID degradation into a Distribution Penalty and a Geometry Penalty, giving a common language for comparing ID quantization methods.
    \item \textbf{Proof-of-Concept Algorithm}: Guided by the framework, we present DRQ as a simple implementation. It decouples continuous geometry learning with a VAE from discrete distribution matching with K-Means, making the two objectives easier to inspect separately.
    \item \textbf{Empirical Validation}: Experiments on a large-scale industrial dataset support the diagnostic value of the framework. The results reveal a trade-off among overlap-based symbolic capacity, codebook utilization, reconstruction fidelity, embedding-level retrieval retention, and behavior-aware soft matching.
\end{enumerate}

\section{Related Work}

\subsection{Semantic IDs in Recommendation}
Early recommendation systems relied on one-hot encodings or hashing. More recent work introduced Semantic IDs as structured discrete representations for retrieval, compression, and unified item modeling \cite{rajput2023recommender, serec2024, onerec2024}. Systems such as TIGER, SE-REC, and OneRec show that discrete item codes can connect multimodal content with collaborative signals while reducing item sparsity. These systems usually treat the tokenizer as a black box. As a result, they do not directly measure how tokenizer geometry and code usage affect robustness and downstream retrieval quality.

\subsection{VQ Methods}
VQ-VAE \cite{vandenoord2017neural} first introduced discrete latent variables into neural representation learning. PQ \cite{jegou2010product} and RQ later became standard paradigms for vector discretization, with RQ-VAE \cite{zeghidour2021soundstream} as a representative neural implementation. We focus on RQ-VAE rather than PQ because RQ-VAE preserves a coarse-to-fine hierarchy that is useful for hierarchical semantic codes, whereas PQ splits vectors into orthogonal subspaces and can weaken global semantic correlation. In joint-training frameworks, however, quantization truncation through the Straight-Through Estimator (STE) couples codebook learning with rigid reconstruction targets. This coupling can distort the continuous latent manifold and trigger index collapse \cite{jing2021understanding}. Recent techniques help with code distribution, but they do not fully address the geometric mismatch created when curved manifolds are forced onto rigid grids. Decoupling representation learning from discrete assignment through clustering \cite{lloyd1982least} offers one way to reduce this failure mode.

\section{Theoretical Analysis}

\subsection{VQ Mapping Dilemma}
In classical VQ paradigms like PQ and neural RQ variants such as RQ-VAE, the model discretizes vectors with one or more codebooks. In a general $L$-level setting, each level $l \in \{1, \dots, L\}$ utilizes a codebook $\mathcal{C}^{(l)} = \{c_1^{(l)}, \dots, c_K^{(l)}\}$ of size $K$.

The quantization process assigns $x \in \mathbb{R}^d$ to a discrete token sequence by finding the nearest centroids across levels, such as matching sub-vectors in PQ or matching residuals in RQ-VAE. The reconstructed vector $\hat{x}$ is typically a linear concatenation in PQ or a linear superposition in RQ-VAE. These mappings assume that the latent space can be partitioned by flat Euclidean cells.

\subsection{Overlap Rate and Effective Codebook Capacity}
In a quantized retrieval system, a small perturbation in the first semantic code level can redirect an item toward a different discrete neighborhood. We call this phenomenon \textit{retrieval-time perturbation}. To quantify the risk, suppose the retrieval pipeline introduces isotropic Gaussian noise with variance $\sigma^2$ when predicting latent vectors. We retain the name \textbf{Overlap Rate ($O_\pi$)}, but mathematically it is a \textit{normalized overlap proxy} rather than an exact Voronoi-boundary crossing probability. It measures how strongly perturbed codeword densities overlap in expectation under the empirical codeword prior. This isotropic perturbation mirrors the K-Means assumption of spherical clusters with equal variances, a premise known to oversimplify complex multimodal distributions \cite{tong2026rqgmm} but also common in RQ-VAE-style vector assignments. For a given codebook, under this spherical equal-variance assumption, $O_\pi$ is defined as (detailed derivation is provided in the Appendix):
\begin{equation}
    O_\pi \approx \sum_{i=1}^K \sum_{j=1}^K \pi_i \pi_j \exp\left(-\frac{\|c_i - c_j\|^2}{4\sigma^2}\right)
\end{equation}
where $\pi_i$ is the prior usage probability of codeword $i$, and $c_i$ is its geometric center. 

Furthermore, we define the \textbf{Effective Codebook Size ($K_{\text{eff}}$)} as $K_{\text{eff}} = 1 / O_\pi$. The definition follows from an ideal case: in a perfectly uniform codebook with infinitely distant codewords ($\pi_i = 1/K$), the off-diagonal overlap vanishes and the remaining diagonal mass is exactly $1/K$. Taking the inverse $1/O_\pi$ therefore estimates how many uniformly used, non-overlapping codewords would produce the same normalized overlap score. A larger $K_{\text{eff}}$ means that a Semantic ID has more effective symbolic capacity without requiring longer token sequences.

Equation (1) reveals the two root causes of Semantic ID degradation:
\begin{itemize}
    \item \textbf{Distribution Penalty}: If codebook collapse occurs, most samples are assigned to a few popular codewords, skewing $\pi_i$. Splitting Equation (1) into diagonal and off-diagonal parts yields a distribution-controlled lower bound
    \begin{equation*}
        O_\pi = \underbrace{\sum_i \pi_i^2}_{\text{distribution floor}} + \underbrace{\sum_{i \neq j} \pi_i \pi_j \exp\left(-\frac{\|c_i - c_j\|^2}{4\sigma^2}\right)}_{\text{geometry-sensitive cross-overlap}}.
    \end{equation*}
    The first term is the self-collision mass: even with perfect geometric separation, a highly concentrated prior keeps $O_\pi$ large. By flattening the usage distribution, we can reduce this unavoidable floor and also downweight off-diagonal overlaps.
    \item \textbf{Geometry Penalty}: When the data lie near a curved manifold (e.g., a hypersphere), the multi-layer Euclidean superposition in RQ-VAE can place reconstructed combinations away from that manifold. Many nominal code combinations may then occupy regions with few or no real items, while real items are concentrated in a smaller portion of the code space. Because the inter-codeword distance $\|c_i - c_j\|^2$ remains small, the cross-overlap term stays large, increasing $O_\pi$ and shrinking $K_{\text{eff}}$. Expanding the latent space and increasing geometric separation can help reduce this mismatch.
\end{itemize}

\section{Methodology}
As established in Section 3, Equation (1) couples two ingredients: the usage weights $\pi_i \pi_j$ and the geometry kernel $\exp(-\|c_i - c_j\|^2/4\sigma^2)$. Reducing $O_\pi$ therefore requires two changes at the same time. The codeword prior should be less concentrated, which lowers the distribution floor $\sum_i \pi_i^2$, and the codewords should be farther apart, which suppresses off-diagonal cross-overlap.

Standard end-to-end quantization makes these goals difficult to satisfy jointly. Gradients flow only through activated codewords via STE. For long-tailed data, popular items dominate these updates, pulling centroids toward dense regions and leaving many codes under-trained. This raises the self-collision floor. At the same time, forcing continuous representations onto a rigid grid limits how far embeddings can spread apart \cite{arvanitidis2018latent}, keeping the geometry kernel large.

DRQ makes this conflict easier to study by separating the two objectives into two stages. First, a continuous VAE learns a latent representation without discrete codebook constraints. The latent embeddings can spread under reconstruction learning, and an optional contrastive loss can add collaborative supervision when behavior data are available. Second, post-hoc K-Means converts the frozen continuous vectors into discrete codes. This clustering stage does not guarantee perfectly uniform code usage, but it replaces sparse codebook updates with dataset-level assignment and centroid recomputation.

\subsection{Continuous Reshaping}
In traditional VQ-VAE, the quantization layer forces the latent space to follow Euclidean grid slicing, which can trigger the Geometry Penalty. DRQ therefore removes the discrete quantization operation from the training loop.

The training of the encoder $E_\phi$ and decoder $D_\theta$ is driven by the reconstruction loss $L_{\text{recon}}$ plus an optional contrastive term $L_{\text{contrastive}}$:
\begin{equation}
    L = \mathbb{E}_{x \sim \mathcal{D}} \left[ \|x - D_\theta(E_\phi(x))\|_2^2 \right] + \lambda L_{\text{contrastive}}
\end{equation}
where $L_{\text{contrastive}}$ is an InfoNCE loss that pulls co-viewed or semantically similar items closer in the latent space. In the base DRQ-VAE experiments, we set $\lambda = 0$ to isolate the effect of decoupling. DRQ-VAE+CL activates this term to inject user-behavior supervision. Without Commitment Loss and STE gradients, the continuous latent variable $z = E_\phi(x)$ can expand before being discretized. When enabled, contrastive learning optimizes two commonly studied properties \cite{wang2020understanding}: (1) \textbf{Alignment}, which clusters similar items, and (2) \textbf{Uniformity}, which pushes random negatives apart on the hypersphere. Uniformity helps enlarge inter-item distance $\|c_i - c_j\|^2$ globally, while alignment forms behavior-aware neighborhoods locally.

\subsection{Discrete Matching}
After the continuous representation is trained and frozen, we apply hierarchical K-Means (RQ-KMeans) to the vectors $\mathcal{Z} = \{z_1, z_2, \dots, z_N\}$ to obtain hierarchical Semantic IDs:
\begin{enumerate}
    \item \textbf{Level 1}: Execute K-Means on the global dataset $\mathcal{Z}$ to find $K$ centroids $\mathcal{C}^{(1)}$. Each point $z_n$ gets label $l_n^{(1)}$, and we compute the residual $e_n^{(1)} = z_n - c_{l_n^{(1)}}^{(1)}$.
    \item \textbf{Level $l$}: Execute K-Means on the residuals $\{e_n^{(l-1)}\}$ to find layer $l$ centroids $\mathcal{C}^{(l)}$ and labels $l_n^{(l)}$, recursively calculating the next residuals.
\end{enumerate}
K-Means adapts its centroids to the observed data density. In long-tailed settings, it allocates more centroids to dense regions and fewer to sparse regions. Because it uses dataset-level assignment and centroid recomputation rather than STE updates on only the currently activated codes, it is less exposed to the update starvation seen in joint VQ training. This does not imply that every code remains active after convergence; deep residual levels can still leave some codes unused. In practice, however, the clustering stage tends to broaden code usage and reduce the distribution floor relative to heavily collapsed codebooks.

\subsection{Role of Continuous Representation}
Because density-adaptive clustering helps mitigate the Distribution Penalty, one might bypass the VAE entirely and apply hierarchical clustering directly to raw item embeddings. This alternative is useful as a baseline, but it can only use the geometry and semantics already present in the input features.

Raw multimodal representations often suffer from representation degeneration and anisotropy (the "cone effect") \cite{gao2018representation, ethayarajh2020contextual}, where items cluster densely in a narrow region. Directly clustering such embeddings can preserve useful structure, but it cannot reshape the representation space or add missing collaborative signals. Raw embeddings also mainly represent static content, such as visual or textual features, and may not contain the behavior information needed for recommendation tasks.

The VAE stage provides \textbf{structural flexibility}. It can fuse heterogeneous multimodal features, such as text, image, and categorical tags, and it can incorporate user behavior signals through sequence or graph-based contrastive learning. It can also accommodate regularization losses, including Information Bottleneck \cite{tishby2015deep} or uniformity constraints \cite{wang2020understanding}. In this view, the continuous reshaping stage is not guaranteed to improve every diagnostic, but it provides a trainable space in which reconstruction fidelity, collaborative semantics, and geometric separation can be adjusted before discrete clustering is applied.

\section{Experiments}
\subsection{Setup}
We evaluate our method on a real-world industrial short-video dataset containing over 15 million items. The initial multimodal dense embeddings have dimension $d=256$. The dataset is proprietary and privacy-sensitive, so the experiments should be interpreted as an industrial case study rather than a cross-benchmark leaderboard.

\textbf{Baseline Methods}: We compare DRQ-VAE with four reference methods, each isolating a different design choice.
\begin{itemize}
    \item \textbf{Classical RQ-VAE}: a standard hierarchical quantizer for Semantic IDs, included as the joint-training baseline.
    \item \textbf{RQP-VAE}: an enhanced RQ-VAE with Exponential Moving Average (EMA) codebook updates and dead-code revival, used to test how much distribution-side stabilization helps.
    \item \textbf{RQ-KMeans}: hierarchical K-Means directly applied to raw input embeddings, used to test what the VAE reshaping stage contributes beyond density-adaptive clustering.
    \item \textbf{DRQ-VAE+CL}: DRQ-VAE with contrastive learning, used to test whether user-behavior supervision improves downstream matching.
\end{itemize}
All methods use a hierarchical codebook of size $K=4096$ and $L=3$ levels, corresponding to a theoretical capacity of 36 bits. In the standard \textbf{DRQ-VAE} evaluation, we set the contrastive weight $\lambda = 0$ to isolate the effect of decoupling. For \textbf{DRQ-VAE+CL}, we set $\lambda = 0.3$ to activate user behavior alignment. Sections 5.2--5.4 report representation diagnostics, while Section 5.5 evaluates item-to-item retrieval built from user behavior sequences.

\subsection{Geometry}
We first evaluate how each tokenizer changes the geometry of the embedding space. The goal is not only to measure reconstruction quality, but also to check whether the latent space remains sufficiently high-dimensional and well spread. We use four metrics:
\begin{itemize}
    \item \textbf{Participation Ratio}: Measures the effective number of dimensions utilized, calculated as $(\text{Tr}(\Sigma))^2 / \text{Tr}(\Sigma^2)$, where $\Sigma$ is the covariance matrix.
    \item \textbf{Entropy-Based Effective Rank}: Defined as $\exp(H(p))$, where $p_i=\lambda_i / \sum_j \lambda_j$ are the normalized eigenvalues of $\Sigma$ and $H(p)=-\sum_i p_i \log p_i$. This metric evaluates how many dimensions effectively carry variance.
    \item \textbf{$\lambda_{\text{max}}$ (Max Eigenvalue)}: The maximum eigenvalue of $\Sigma$ (normalized), indicating the proportion of variance captured by the first principal component.
    \item \textbf{Mean Abs Cosine}: Average absolute cosine similarity between all pairs of latent dimensions, assessing the overall orthogonality of the space.
\end{itemize}
Table \ref{tab:geometry} shows these geometric statistics calculated on a 20,000 random subset.

\begin{table}[h]
    \centering
    \caption{Comparison of Latent Geometry and Topology for Semantic ID Construction}
    \label{tab:geometry}
    \resizebox{\columnwidth}{!}{%
    \begin{tabular}{lcccc}
        \toprule
        \textbf{Model} & \textbf{Participation Ratio} & \textbf{Entropy-Based Effective Rank} & \textbf{$\lambda_{\text{max}}$} & \textbf{Mean Abs Cosine} \\
        \midrule
        Raw Input & 106.05 & 131.17 & 0.0307 & 0.0795 \\
        RQ-VAE & 73.08 & 116.77 & 0.0456 & 0.0865 \\
        RQP-VAE & 104.91 & 137.77 & 0.0313 & \textbf{0.0720} \\
        RQ-KMeans & 105.92 & 131.15 & 0.0309 & 0.0752 \\
        DRQ-VAE & 71.79 & 103.01 & 0.0400 & 0.0894 \\
        DRQ-VAE+CL & \textbf{149.19} & \textbf{189.54} & \textbf{0.0159} & 0.4622 \\
        \bottomrule
    \end{tabular}%
    }
\end{table}

\textbf{Analysis}: For Semantic IDs to remain useful in recommendation and retrieval, the latent space should preserve many discriminative directions without becoming overly anisotropic. Under the current $L=3$, $K=4096$ setting, RQ-VAE still degrades geometry relative to the raw input: its Participation Ratio drops from 106.05 to 73.08, its entropy-based effective rank drops from 131.17 to 116.77, and both $\lambda_{\text{max}}$ and Mean Abs Cosine increase. The degradation is moderate rather than a near-one-dimensional collapse.

RQP-VAE and RQ-KMeans stay close to the raw geometry. RQP-VAE achieves the lowest Mean Abs Cosine (0.0720), indicating the cleanest near-orthogonal latent organization among the quantized models. DRQ-VAE does not dominate these geometry metrics in this configuration, although it later performs strongly on reconstruction. DRQ-VAE+CL shows the sharpest trade-off: it achieves the best Participation Ratio (149.19), the best entropy-based effective rank (189.54), and the lowest $\lambda_{\text{max}}$ (0.0159), but its Mean Abs Cosine rises to 0.4622. This suggests that contrastive supervision spreads variance across more directions while also introducing stronger inter-dimension correlations.

\subsection{Robustness \& Capacity}
To validate Equation (1), we inject standard Gaussian noise scaled by the empirical latent variance ($\sigma^2 = 1.0 \times \text{Var}(Z)$) into the latent space to test robustness. Results are in Table \ref{tab:overlap}.

\begin{table}[h]
    \centering
    \caption{Expected Overlap and Effective Capacity under Retrieval-Time Perturbation}
    \label{tab:overlap}
    \resizebox{\columnwidth}{!}{%
    \begin{tabular}{l|cc|cc|cc}
        \toprule
        & \multicolumn{2}{c|}{\textbf{Level 0}} & \multicolumn{2}{c|}{\textbf{Level 1}} & \multicolumn{2}{c}{\textbf{Level 2}} \\
        \textbf{Model} & $O_\pi$ & $K_{\text{eff}}$ & $O_\pi$ & $K_{\text{eff}}$ & $O_\pi$ & $K_{\text{eff}}$ \\
        \midrule
        RQ-VAE & 0.002196 & 455.46 & 0.000466 & 2146.12 & 0.000427 & 2344.55 \\
        RQP-VAE & \textbf{0.000273} & \textbf{3667.92} & \textbf{0.000314} & \textbf{3180.69} & \textbf{0.000310} & \textbf{3221.32} \\
        RQ-KMeans & 0.000285 & 3506.33 & 0.000575 & 1738.18 & 0.000644 & 1553.27 \\
        DRQ-VAE & 0.000286 & 3495.31 & 0.000512 & 1951.97 & 0.000549 & 1821.62 \\
        DRQ-VAE+CL & 0.000297 & 3365.96 & 0.001075 & 930.04 & 0.001962 & 509.73 \\
        \bottomrule
    \end{tabular}%
    }
\end{table}

\textbf{Analysis}: Table \ref{tab:overlap} reports the $\sigma^2 = \mathrm{Var}(Z)$ slice of the perturbation analysis and estimates how many non-confusable codewords each method provides. RQP-VAE performs best on this proxy. It has the lowest $O_\pi$ and the largest $K_{\text{eff}}$ at all three levels, showing that EMA-based distribution flattening is highly effective for symbolic robustness under isotropic noise.

RQ-VAE is the weakest baseline overall, but its main failure appears at Level 0 rather than at the deepest level. Its coarse codebook has $O_\pi=0.002196$ and only 455.46 effective states, while the deeper levels recover to more than 2,100 effective states. DRQ-VAE and RQ-KMeans occupy the middle range, with DRQ-VAE slightly better than RQ-KMeans at Levels 1 and 2 but still behind RQP-VAE. DRQ-VAE+CL degrades in the deeper codebooks: its $K_{\text{eff}}$ drops from 3365.96 at Level 0 to 930.04 at Level 1 and 509.73 at Level 2. Collaborative reshaping therefore helps downstream soft matching, but it does not automatically improve overlap-based symbolic robustness.

\subsection{Codebook Utilization}
We examine codebook usage across the entire dataset using three metrics: \textbf{Perplexity} ($\exp(H(p))$, where $H(p)$ is the Shannon entropy of codeword frequencies, measuring effective codebook size up to $K$), \textbf{Active Codes} (number of codewords assigned to at least one item), and \textbf{Gini} (Gini coefficient of codeword frequencies, measuring distribution inequality). Results are in Table \ref{tab:codebook}.

\begin{table}[h]
    \centering
    \caption{Codebook Utilization Statistics for Semantic ID Construction (Max 4096)}
    \label{tab:codebook}
    \resizebox{\columnwidth}{!}{%
    \begin{tabular}{clccc}
        \toprule
        \textbf{Level} & \textbf{Model} & \textbf{Perplexity} & \textbf{Active Codes} & \textbf{Gini} \\
        \midrule
        \multirow{5}{*}{L0} 
        & RQ-VAE & 480.96 & 512 & 0.899 \\
        & RQP-VAE & \textbf{3872.12} & \textbf{4096} & \textbf{0.184} \\
        & RQ-KMeans & 3765.45 & \textbf{4096} & 0.225 \\
        & DRQ-VAE & 3758.63 & \textbf{4096} & 0.229 \\
        & DRQ-VAE+CL & 3665.12 & \textbf{4096} & 0.258 \\
        \midrule
        \multirow{5}{*}{L2} 
        & RQ-VAE & 2815.64 & 3377 & 0.445 \\
        & RQP-VAE & \textbf{3692.76} & \textbf{4096} & \textbf{0.239} \\
        & RQ-KMeans & 1912.13 & 3371 & 0.636 \\
        & DRQ-VAE & 2130.85 & 3464 & 0.592 \\
        & DRQ-VAE+CL & 698.71 & 2389 & 0.865 \\
        \bottomrule
    \end{tabular}%
    }
\end{table}

\textbf{Analysis}: These statistics show whether the nominal 4096-way codebooks become distinct semantic neighborhoods in practice. RQ-VAE's most severe collapse occurs at Level 0: its perplexity falls to 480.96, only 512 codes are active, and the Gini coefficient reaches 0.899. This indicates a highly concentrated coarse partition.

RQP-VAE is the strongest model on pure utilization at both the coarse and fine levels, reaching the highest perplexity, full code activation, and the lowest inequality at L0 and L2. RQ-KMeans and DRQ-VAE both retain moderate fine-level usage without explicit dead-code revival, with DRQ-VAE slightly improving over RQ-KMeans at L2 through higher perplexity and more active codes. DRQ-VAE+CL moves in the opposite direction: its deepest codebook utilization deteriorates, with perplexity 698.71, only 2389 active codes, and Gini 0.865 at L2. In this run, collaborative supervision helps soft downstream matching more than it helps deep discrete code usage.

\subsection{Item-to-Item Retrieval Performance}
To connect the representation diagnostics above with a downstream task, we evaluate item-to-item retrieval from user behavior sequences. User sequences are stored in newest-first order. For every consecutive pair $(\text{query}, \text{target})=(\text{older item}, \text{newer item})$, we create one test case if both items appear in the evaluation pool. Otherwise, the pair is skipped, and we do not perform gap jumping. Retrieval is carried out over the full item pool with the query item removed from the candidate set.

\begin{table}[h]
    \centering
    \caption{Item-to-Item Retrieval Retention under SID-Reconstructed Embeddings.}
    \label{tab:i2i_retention}
    \resizebox{\columnwidth}{!}{%
    \begin{tabular}{lcccc}
        \toprule
        \textbf{Model} & \textbf{@20} & \textbf{@50} & \textbf{@100} & \textbf{@200} \\
        \midrule
        RQ-VAE & 0.5561 & 0.6472 & 0.7162 & 0.7707 \\
        RQP-VAE & 0.6625 & 0.7365 & 0.7984 & 0.8551 \\
        RQ-KMeans & 0.5589 & 0.6297 & 0.6985 & 0.7765 \\
        DRQ-VAE & \textbf{0.9999} & \textbf{0.9998} & 0.9999 & 0.9997 \\
        DRQ-VAE+CL & 0.9976 & 0.9991 & \textbf{1.0016} & \textbf{1.0045} \\
        \bottomrule
    \end{tabular}%
    }
\end{table}

\begin{table}[h]
    \centering
    \caption{Item-to-Item Retrieval AUC Metrics.}
    \label{tab:i2i_auc}
    \resizebox{\columnwidth}{!}{%
    \begin{tabular}{lccc}
        \toprule
        \textbf{Model} & \textbf{SID Embedding AUC} & \textbf{Weighted SID Match AUC} & \textbf{Exact SID Match AUC} \\
        \midrule
        RQ-VAE & 0.9112 & 0.9127 & \textbf{0.7946} \\
        RQP-VAE & 0.9112 & 0.8907 & 0.7579 \\
        RQ-KMeans & 0.9103 & 0.8872 & 0.7527 \\
        DRQ-VAE & 0.9114 & 0.9016 & 0.7466 \\
        DRQ-VAE+CL & \textbf{0.9121} & \textbf{0.9240} & 0.7553 \\
        \bottomrule
    \end{tabular}%
    }
\end{table}

\textbf{Protocol Details}: The evaluation reports three groups of metrics. First, it retrieves neighbors using the original embedding, the SID-reconstructed embedding, and a random baseline, then converts the reconstructed result into a retention score $\mathrm{HR}_{\text{sid}}/\mathrm{HR}_{\text{orig}}$. This group measures how much geometry is preserved after quantization and reconstruction. Second, it performs discrete SID retrieval. Hierarchical models use prefix matching, while non-hierarchical models use independent code matching. These detailed HR@K statistics are reported in the evaluation log and are most sensitive to code collisions and distribution flatness. Third, it computes three AUC metrics on the same sliding-window test pairs with fixed random negatives: \textbf{SID Embedding AUC}, \textbf{Weighted SID Match AUC}, and \textbf{Exact SID Match AUC}. These metrics distinguish soft semantic similarity from strict symbolic equality. We omit the original upper bound, random baseline, and duplicate-count diagnostics from the main tables to keep the presentation compact.

\textbf{Analysis}: Tables \ref{tab:i2i_retention} and \ref{tab:i2i_auc} show a split between reconstruction fidelity, soft semantic matching, and exact symbolic lookup. DRQ-VAE is the strongest reconstruction model by a wide margin. The full evaluation log reports the lowest MSE (0.000432), the highest cosine similarity (0.999784), the lowest collision rate (0.074680), the largest number of unique IDs (462,660), and the smallest maximum collision bucket (35). Consistent with these reconstruction statistics, DRQ-VAE delivers nearly lossless retention at the lower cutoffs, reaching 0.9999 at @20 and 0.9998 at @50.

DRQ-VAE+CL becomes strongest once retrieval depends more heavily on soft semantic similarity or broader candidate sets. It achieves the best high-cutoff retention at @100 and @200 (1.0016 and 1.0045), as well as the best \textbf{SID Embedding AUC} (0.9121) and \textbf{Weighted SID Match AUC} (0.9240). This indicates that collaborative supervision is most useful when retrieval is evaluated through softly weighted semantic matching rather than reconstruction fidelity alone.

Exact symbolic lookup follows a different pattern. \textbf{Exact SID Match AUC} is highest for RQ-VAE (0.7946), not necessarily because it has the cleanest discrete partition, but partly because it reuses codes more aggressively: its collision rate is 0.228898, it yields only 385,551 unique IDs, and its largest collision bucket reaches 76. These statistics suggest that stronger exact-match scores can arise from heavier code sharing rather than cleaner semantic separation. Taken together, the downstream evidence supports a three-way trade-off: RQP-VAE is strongest on symbolic capacity proxies, DRQ-VAE is strongest on reconstruction fidelity and near-lossless embedding retrieval retention, and DRQ-VAE+CL is strongest on soft matching and high-cutoff retrieval.

\subsection{Discussion and Limitations}
The current empirical evidence should be interpreted with several caveats. First, all experiments are conducted on a single proprietary industrial short-video dataset. This setting is useful for stress-testing large-scale Semantic ID construction, but it does not establish that the same ranking among DRQ, RQP-VAE, and RQ-KMeans will hold on public recommendation benchmarks with different sparsity patterns, item modalities, or sequence dynamics.

Second, Section 5.5 studies item-to-item retrieval rather than full recommendation pipelines such as candidate generation followed by ranking. The downstream claims are therefore about the quality of Semantic IDs as retrieval and matching keys, not about end-to-end recommendation metrics such as Recall or NDCG in a production stack.

Third, the proposed diagnostics are simplified proxies: $O_\pi$ assumes isotropic perturbation, and the retrieval metrics separate reconstructed, weighted, and exact matching views instead of collapsing them into a single score. This decomposition is intentional, but broader validation is still needed. Future work should evaluate the framework on public datasets, additional downstream tasks, and more diverse Semantic ID tokenizers.

\section{Conclusion}
This paper develops a quantitative framework for diagnosing Semantic ID degradation in recommendation systems. By formalizing $O_\pi$ and $K_{\text{eff}}$, the framework separates two failure sources: distribution skew in code usage and geometric overlap among codewords. DRQ serves as a proof-of-concept implementation of this view by separating continuous representation learning from discrete code assignment, but the experiments show that this decoupling should be interpreted as a controllable design choice rather than a uniformly superior tokenizer. Under the current three-level, 4096-codeword setting, no single tokenizer dominates every view of Semantic ID quality. RQP-VAE is strongest on overlap-based capacity and codebook utilization, DRQ-VAE is strongest on reconstruction fidelity and near-lossless embedding retrieval retention, and DRQ-VAE+CL is strongest on soft Semantic-ID matching and high-cutoff retrieval retention. The central conclusion is therefore that geometry-sensitive retrieval, symbolic robustness, and behavior-aware soft matching are related but distinct objectives. They should not be collapsed into a single notion of tokenizer quality.

\clearpage
\appendix

\section*{APPENDIX}

\section{Derivation of the Overlap Proxy $O_\pi$ and Effective Capacity $K_{\text{eff}}$}

To quantify the robustness of Semantic IDs under retrieval-time perturbation, we derive the normalized overlap proxy $O_\pi$ induced by Gaussian uncertainty around codewords.

\textbf{Assumptions and Definitions:} Let the continuous latent space be $\mathbb{R}^d$ and the codebook be $\mathcal{C} = \{c_1, \dots, c_K\}$. We assume that the retrieval system induces isotropic Gaussian uncertainty around each codeword $c_i$, represented by the probability density function $p_i(x) = \mathcal{N}(x | c_i, \sigma^2 I_d)$.

\textbf{Pairwise Overlap Integral:} We define the overlap between two codewords $c_i$ and $c_j$ as the integral of the product of their density functions:
\begin{equation}
\text{overlap}_{ij} \triangleq \int p_i(x) p_j(x) dx
\end{equation}
Expanding the product of the two Gaussians yields:
\begin{equation}
p_i(x)p_j(x) = \frac{1}{(2\pi\sigma^2)^d} \exp\left( - \frac{\|x - c_i\|^2 + \|x - c_j\|^2}{2\sigma^2} \right)
\end{equation}
By completing the square and introducing the midpoint $\mu_{ij} = (c_i + c_j)/2$, we can rewrite the exponent numerator:
\begin{equation}
\|x - c_i\|^2 + \|x - c_j\|^2 = 2\|x - \mu_{ij}\|^2 + \frac{1}{2}\|c_i - c_j\|^2
\end{equation}
Substituting this back, the product factors into a term depending on $x$ and a constant term:
\begin{equation}
p_i(x)p_j(x) = \frac{1}{(2\pi\sigma^2)^d} \exp\left( - \frac{\|x - \mu_{ij}\|^2}{\sigma^2} \right) \exp\left( - \frac{\|c_i - c_j\|^2}{4\sigma^2} \right)
\end{equation}
Integrating over all $x \in \mathbb{R}^d$, the first exponential term evaluates to $(\pi\sigma^2)^{d/2}$. Thus, the exact pairwise overlap is:
\begin{equation}
\text{overlap}_{ij} = \frac{1}{(4\pi\sigma^2)^{d/2}} \exp\left( - \frac{\|c_i - c_j\|^2}{4\sigma^2} \right)
\end{equation}

\textbf{Normalized Overlap Kernel:} The diagonal self-overlap is
\begin{equation}
\text{overlap}_{ii} = \frac{1}{(4\pi\sigma^2)^{d/2}}.
\end{equation}
Dividing by this common self-overlap yields a scale-free kernel
\begin{equation}
\mathcal{K}(c_i, c_j) = \frac{\text{overlap}_{ij}}{\text{overlap}_{ii}} = \exp\left( - \frac{\|c_i - c_j\|^2}{4\sigma^2} \right),
\end{equation}
which equals 1 on the diagonal and decays toward 0 as codewords separate.

\textbf{Global Expected Overlap Proxy $O_\pi$:} Let $\pi_i$ be the prior probability of utilizing codeword $c_i$ ($\sum \pi_i = 1$). We define $O_\pi$ as the expectation of this normalized overlap kernel under the empirical codeword prior:
\begin{equation}
O_\pi = \sum_{i=1}^K \sum_{j=1}^K \pi_i \pi_j \exp\left( - \frac{\|c_i - c_j\|^2}{4\sigma^2} \right)
\end{equation}
This quantity is not an exact Voronoi-boundary crossing probability. Instead, it is a normalized confusion proxy: lower values mean that perturbed codeword densities overlap less in expectation.

\textbf{Effective Codebook Size $K_{\text{eff}}$:} Under ideal uniform and orthogonal conditions ($\pi_i = 1/K$ and infinite distance), the off-diagonal kernel vanishes and the overlap lower bound is:
\begin{equation}
O_{\pi, \text{uniform}} = \sum_{i=1}^K \left(\frac{1}{K}\right)^2 = \frac{1}{K}
\end{equation}
To express a measured overlap as the equivalent number of codewords in a perfectly uniform system, we define Effective Codebook Size as:
\begin{equation}
K_{\text{eff}} = \frac{1}{O_\pi}
\end{equation}
Thus, maximizing $K_{\text{eff}}$ requires both lower distribution skewness (flatter $\pi$) and larger geometric separation $\|c_i - c_j\|^2$. This is the mathematical foundation for the decoupled approach.

\bibliographystyle{unsrt}
\bibliography{references}

\end{document}